\documentclass[a4paper, 10pt, oneside]{article}
\def\Draft{1}
\def\Draft{0}
\def\Supp{0}
\def\Supp{1}
\pdfoutput=1
\newcommand{\Author}{Perrinet}%
\newcommand{\FirstName}{Laurent U}%
\newcommand{\Institute}{Institut de Neurosciences de la Timone (UMR7289),\\ CNRS / Aix-Marseille Universit\'e}%
\newcommand{\Address}{27, Bd. Jean Moulin, 13385 Marseille Cedex 5, France}%
\newcommand{\Website}{https://laurentperrinet.github.io/}%
\newcommand{\Email}{laurent.perrinet@univ-amu.fr}%
\newcommand{\Title}{Differential response of the retinal neural code with respect to the spatial sparseness of natural images}%
\newcommand{\Abstract}{
Natural images follow statistics inherited by the structure of our physical (visual) environment. In particular, a prominent facet of this structure is that images can be described by a relatively sparse number of features. Strikingly, this sparseness has a strong spatial component such that in general some sub-parts of an image are  uniform while others contain clusters of textural (dense) structures. To investigate the role of this facet of images' sparseness in the efficiency of the neural code, we designed a new class of random textured stimuli with a controlled parameter inspired by measurements made on a dataset of natural images. Then, we tested the impact of this parameter on the firing pattern observed in a population of retinal ganglion cells recorded \emph{ex vivo} in the retina of a diurnal rodent. These recordings showed in particular that the reliability of spike timings co-varied with respect to the spatial sparseness with a similar trend than that observed in natural images. These results suggest that the code represented in the spike pattern of ganglion cells may adapt to this aspect of the statistics of natural images.
}
\newcommand{\Keywords}{Retina, Sparseness, Image texture, Computer vision, Neuroscience}%
\newcommand{\Acknowledgments}{%
C.R.R.\ was supported by CONICYT scholarship 21130863. A.G.P.\ was supported by FONDECYT 1150638,  Millennium Institute ICM- P09-022-F, NICOP N62909-14-1-N121 ONRG. M.-J.E. was supported by CONICYT Basal Project FB0008 and FONDECYT 1140403. L.U.P.\ was supported by ANR projects "BalaV1" (ANR-13-BSV4-0014-02) and "TRAJECTORY" (ANR-15-CE37-0011). Correspondence and requests for materials should be addressed to LUP (email:\Email ). Code and supplementary material available at \url{\Website/publication/ravello-16-droplets}.
} %

\usepackage[utf8]{luainputenc}
\usepackage[english]{babel}%
\usepackage[autostyle]{csquotes}

     \usepackage{textcomp}
     \usepackage{libertine}
     \usepackage[varqu,varl]{inconsolata}
     \usepackage[libertine,bigdelims,vvarbb]{newtxmath} 
     \usepackage[cal=boondoxo]{mathalfa} 
     \usepackage[supstfm=libertinesups,%
       supscaled=1.2,%
       raised=-.13em]{superiors}
\usepackage[pdftex]{graphicx}
\DeclareGraphicsExtensions{.pdf}
\graphicspath{{./figures/}}%
\usepackage[unicode,linkcolor=blue,citecolor=blue,filecolor=black,urlcolor=blue,pdfborder={0 0 0}]{hyperref}%
\hypersetup{%
pdftitle={\Title},%
pdfauthor={\Author < \Email > \Address - \Website },%
pdfkeywords={\Keywords},%
pdfsubject={\Acknowledgments}%
}%
\usepackage{url}            
\usepackage{booktabs}       
\usepackage{amsfonts}       
\usepackage{nicefrac}       
\usepackage{microtype}      

\usepackage{tikz}%
\if 1\Draft
\usepackage{setspace}
\fi

\newcommand{\eql}[1]{\begin{equation}#1\end{equation}}

\newcommand{\NN}{\mathbb{N}}
\newcommand{\RR}{\mathbb{R}}

\usepackage{siunitx}
\newcommand{\ms}{\si{\milli\second}}%
\newcommand{\s}{\si{\second}}%
\usepackage[natbib=true,
			style=nature, 
      		sortcites=true,
			abbreviate=true,
			maxcitenames=2,
			maxnames = 5,
			doi=true,
			url=true,
			isbn=false,
			eprint=false,
			texencoding=utf8,
			bibencoding=latin1,
			backend=bibtex,
			]{biblatex}%
\AtEveryBibitem{
  \clearfield{month}
  \clearfield{day}
  \clearfield{note}
  \clearfield{comment}
}
\title{
\Title
\if\Supp1
{\it (annotated manuscript + supplementary material)}
\fi
}

\author{%
  \makebox[.5\linewidth]{Cesar R.~Ravello}\\ PhD Program in Neuroscience\\
Centro Interdisciplinario de Neurociencia de Valpara\'iso \\
Facultad de Ciencias, Universidad de Valparaiso\\
2360102 Valpara\'iso, Chile\\
  \texttt{cesar.ravello@cinv.cl} \\
  \and \makebox[.5\linewidth]{Maria-Jose Escobar}\\Universidad T\'ecnica Federico Santa Mar\'ia\\
Departamento de Electr\'onica\\
2390123 Valpara\'iso, Chile\\
\texttt{mariajose.escobar@usm.cl} \\
  \and \makebox[.5\linewidth]{Adrian G Palacios}\\Centro Interdisciplinario de Neurociencia de Valpara\'iso \\
Facultad de Ciencias, Universidad de Valparaiso\\
2360102 Valpara\'iso, Chile\\
\texttt{adrian.palacios@uv.cl}\\
  \and \makebox[.5\linewidth]{\FirstName\ \Author\thanks{See {\Website}.}}\\\Institute\ \\ \Address\ \\
\texttt{\Email} \\
}
\date{Monday, November 21, 2016}
\begin{document}
\if\Draft1
\doublespacing
\fi
\maketitle
\section*{Abstract}
\Abstract

\textbf{Keywords} : \Keywords
\paragraph{BibTex entry}~~\\
\begin{verbatim}
@unpublished{Ravello17droplets,
  Title = {Differential response of the retinal neural code 
           with respect to the sparseness of natural images},
  AUTHOR = {Ravello, Cesar U and 
            Escobar, Maria-Jose U and 
            Palacios, Adrian G and 
            Perrinet, Laurent U},
  NOTE = {working paper or preprint},
  YEAR = {2016},
  MONTH = Nov,
  KEYWORDS = { Image texture ;  Neuroscience ;  Computer vision ; Retina},
  URL = {https://laurentperrinet.github.io/publication/ravello-16-droplets/},
}
\end{verbatim}
\if\Draft1
\newpage 
\doublespacing
\fi
\section{Motivation}
\begin{figure}
\centering{
\begin{tikzpicture}
\draw [anchor=north west] (0, 0) node {\includegraphics[width=.31\linewidth]{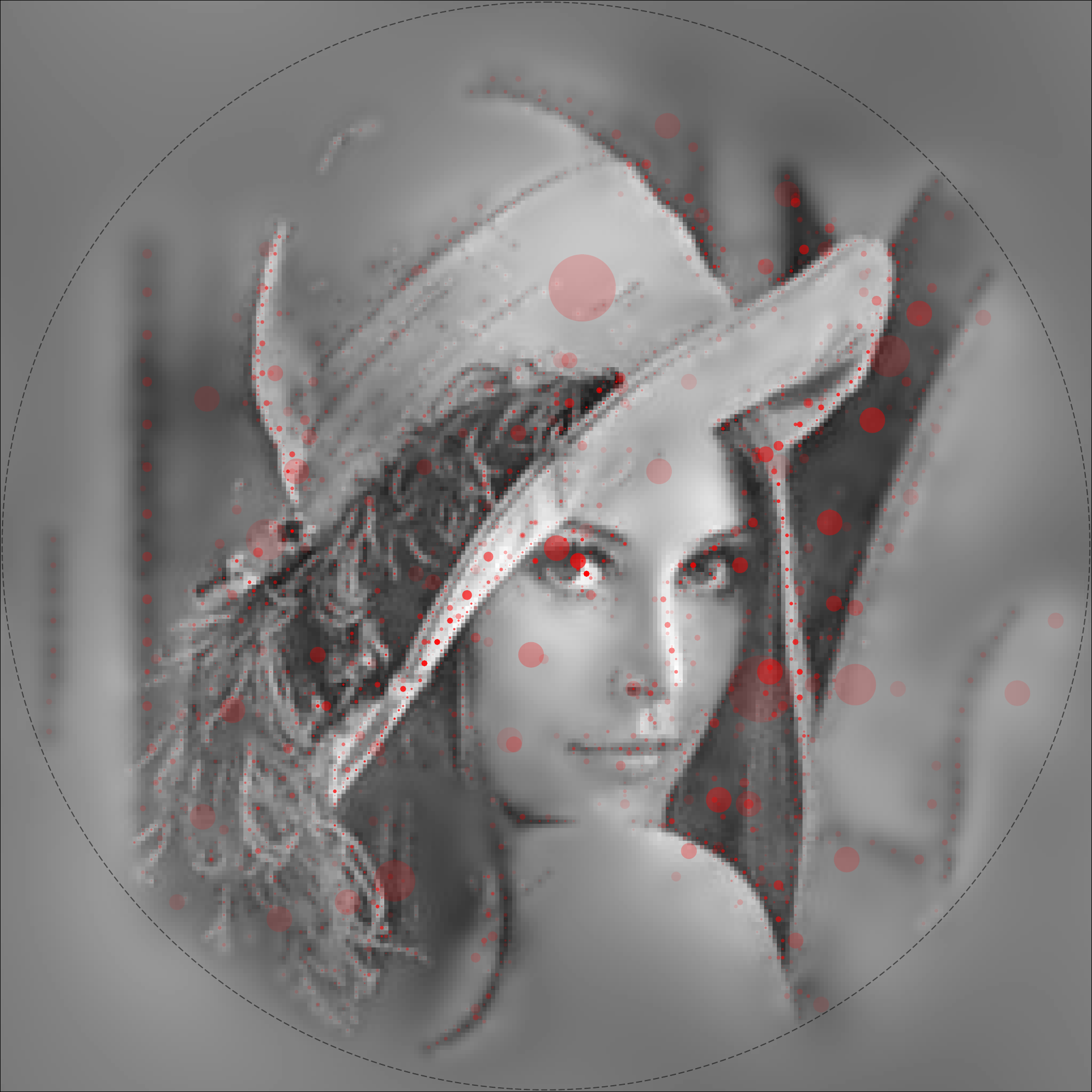}};
\draw [anchor=north west] (.33\linewidth, 0) node {\includegraphics[width=.33\linewidth]{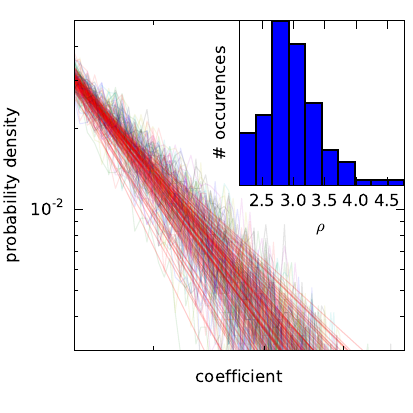}};
\draw [anchor=north west] (.66\linewidth, 0) node {\includegraphics[width=.33\linewidth]{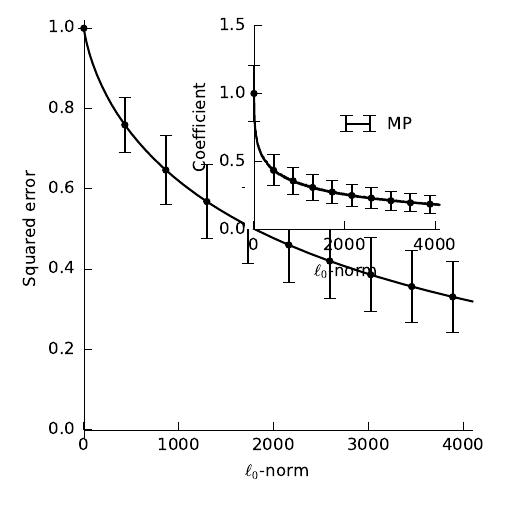}};
\draw (.00\linewidth, 0) node [above right=0mm] {$\mathsf{A}$};
\draw (.33\linewidth, 0) node [above right=0mm] {$\mathsf{B}$};
\draw (.66\linewidth, 0.) node [above right=0mm] {$\mathsf{C}$};
\end{tikzpicture}}
\caption{
{\bf Sparse coding of images in the retina follows regular statistics at the global, not the local scale}: 
\textsf{(A)}~ An instance of the reconstruction of a natural image (``Lena'') with the list of extracted image elements overlaid\if\Supp1\ --- see SI Section~\ref{sec:sparse} for a full description of the algorithm\fi. Edges outside the dashed circle are discarded to avoid artifacts. Parameters for each element are its position, scale and scalar amplitude. \textsf{(B)}~When observing on an image-by-image basis the probability distribution function (pdf) of the coefficients's amplitude, each follows a generic power-law pdf, only controlled by a sparseness parameter $\rho$ for each of the natural images, higher values of $\rho$ denoting more sparse structures. \textsf{(B-inset)}~The distribution of these sparseness parameters shows some variability of values from the most sparsely distributed (right, $\rho\approx4$) to those containing mostly dense textures (left, $\rho\approx2$) but centered around a median vaalue of $\rho\approx3$.
\textsf{(C)}~However, when performing the same analysis locally, that is, on subparts of each image, the pdf of coefficients showed a greater variability. We show here the particular values for different areas of the Lena image: the plume, the face, the background. While the shape of the pdf still follows a power-law, the pdfs are respectively shifted horizontally in the plot. 
This variability is best described by a relative number of coefficients that are greater than a threshold measuring background noise. When merging these results together at the global scale  of the image, these pdfs assemble in a similar power law.
\label{fig:retina_sparseness}}%
\end{figure}
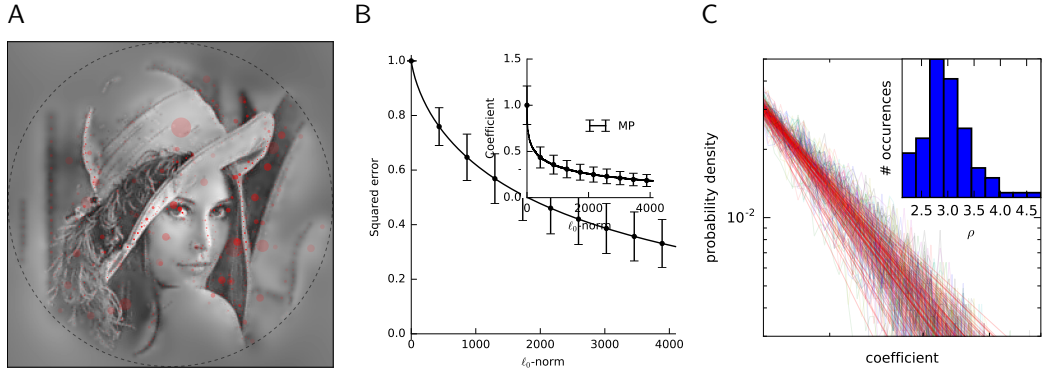%
Natural images, that is, visual scenes that are relevant for an animal, most generally consist of the composition of visual objects, from vast textured backgrounds to single isolated items. By the nature of the structure of the physical (visual) word, these visual items are often sparsely distributed in space, such that these elements are clustered and a large portion of the space is empty: A typical image thus contains vast areas which are containing little information. A simple algorithm to quantify this sparseness is to consider the coding of natural images using a bank of filters resembling the Mexican-hat profiles observed in the retina of most mammals\if\Supp1 (See Supplementary Figure~\ref{fig:retina_sparseness_dog})\fi. Then, it is possible to determine the efficiency of a near-to-optimal coding formalism~\citep{Perrinet15bicv} to show that any image from a database of static, grayscale natural images may be coded solely by a few coefficients (See Figure~\ref{fig:retina_sparseness}\if\Supp1 ; see SI Section~\ref{sec:sparse} for a full description of the algorithm\fi). Moreover, we observed that on this database, some images were sparser than others but that globally they all fitted well a similar power-law probability distribution function. The exponent of this power-law provides an unique descriptor for the sparseness of sparse coefficients and we found that these coefficients were stable across the database~\citep{Perrinet16EUVIP}. However, visual inspection of natural images show that visual information is spatially clustered. Generally, some sub-regions of the image contain dense, textural information (the plumes in Figure~\ref{fig:retina_sparseness}-A and B1) while some are empty (the shoulder in Figure~\ref{fig:retina_sparseness}-A and B2). To date, it is largely unclear how this property inherent to any natural image may be used in the early visual system and in particular in the retina. %

Indeed, a normative explanation for the coding of sensory inflow into spike patterns in the optic nerve is the optimal representation of visual information~\citep{Atick92,Doi2012}. A popular technique to challenge this hypothesis is to test the system using complex stimuli~\citep{Touryan01} and in the retina, this takes the form of the use of stochastic stimuli and the inversion of a model of the neural transformation~\citep{Simoncelli2004}. A major result is that the receptive field of ganglion cells captures information according to maximally informative directions~\citep{Sharpee2004a}. Various aspects of the complex processing that happens in the retina has been captured in models for describing the structure of the retina with respect to spike patterns, from Spike-Triggered average~\citep{Paninski2006}, Spike-Triggered Covariance and quadratic models~\citep{Rajan2013} to more general techniques using variants of the Linear / Non-Linear model~\citep{Simoncelli2004}. However, a difficulty is that these models often assume that the noise models are dense and that introducing sparseness results in non-linear effects which are hard to describe using these models~\citep{Fournier11}. As such, a major challenge is to refine the definition of neural coding so that they conform to the widest variety of results. 

Here, we seek to engender a better understanding of neural coding in the retina by using a novel generative model for dynamic texture synthesis. Our first contribution is to show that while the distribution of sparse coefficients in natural images is stereotyped, it displays a distribution of spatial sparseness parameters (see Figure~\ref{fig:retina_sparseness}-C). From that perspective, we motivate the generation of an optimal stimulation within a stationary Gaussian dynamic texture model. We base our model on a previously defined heuristic coined ``Motion Clouds''~\citep{Leon12}. Our second contribution is an efficient extension of this model which allows to parametrize the sparseness in the texture, for instance based on the measurements made on natural images. 
Our third contribution in this paper is to demonstrate an effect of this sparseness on the spike pattern of retinal ganglion cells from a diurnal rodent, the \emph{Octodon degus} as recorded \emph{ex vivo} on an electrode grid \citep{palacios-etal:14}. Our last contribution is to quantify the efficiency of the neural coding in these recordings and to show that the code is more reliable with a given spatial sparseness level rather than with empty or  dense stimuli. These contributions shows that overall, the processing in the retina may be optimized to match the spatial sparseness encountered in natural scenes. Beyond the contribution to the understanding of the neural code underlying image representation, this may have long ranging applications, for instance for the optimal parametrization of stimulations in a retinal implant.
\section{Design of the ``DropLets'' stimuli}
\begin{figure}
\centering{
\includegraphics[width=.99\linewidth]{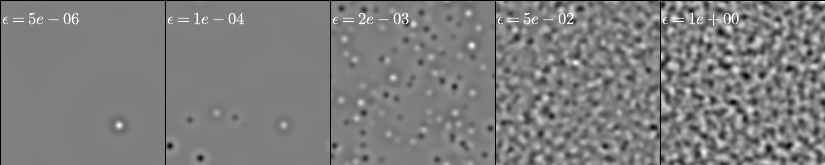}
}
\caption{
{\bf DropLets stimuli:} Based on the Motion Clouds framework, we design DropLets as random phase textures similar to Motion Clouds but where, similarly to the variety observed in a set of natural images, the distribution of coefficients is parameterized by different levels of sparseness.
For the sake of simplicity, sparseness is obtained here by thresholding a mask value and is given by the ratio $\epsilon$ of non-zero coefficients, from very sparse ($\epsilon=5.10^{-6}$) to fully dense ($\epsilon=1=100\%$), and with $\epsilon$ parameters chosen on a geometrical scale.
We show here one single frame of each dynamical stimulus. Note that, while for  $\epsilon=1$ these textures are fully dense and correspond to a linearly filtered noise, the other versions correspond to progressively sparser textures as  $\epsilon$ tends to zero.
Moreover, while the textons used to synthesize these textures are arbitrary, we have chosen here to use symmetrical Mexican-hat profiles (``drops''\if\Supp1 , see Supplementary Figure~\ref{fig:retina_sparseness_dog}\fi) which are known to preferentially evoke activity in the retina.
\label{fig:DropLets}}%
\end{figure}%
From these observations, the understanding of the neural code in the retina is linked to the definition of an adequate generative model. We propose a mathematically-sound derivation of a general parametric model of dynamic textures. This model is defined by aggregation, through summation, of a basic spatial ``texton'' template $\phi(x, y, t)$, eventually transformed by any arbitrary visuo-geometric transformation $g_\lambda$ such as zooms or rotations, which themselves are parameterized by $\lambda$. The summation reflects a transparency hypothesis, which has been adopted for instance in~\citep{Galerne11}. In particular, this simple generative model considers independent, transparent elementary features. While one could argue that this hypothesis is overly simplistic and does not model occlusions or edges, it leads to a tractable framework of stationary Gaussian textures, which has proved useful to model static micro-textures~\citep{Galerne11} and dynamic natural phenomena~\citep{2014-xia-siims}. In particular, the simplicity of this framework allows for a fine tuning of frequency-based (Fourier) parameterization, which is desirable for the interpretation of neuro-physiological experiments. In summary, it states that luminance  $I(x, y, t)$ for $(x, y, t) \in \RR^2 \times \RR$ is  defined as a random field~\citep{Vacher15}:
\eql{\label{eq-deadleaves}
I (x, y, t) = \sum_{i \in \NN} a_i \cdot g_{\lambda_i}(\phi(x-x_i, y-y_i, t-t_i))}
where the respective random variables parameterize respectively each texton's geometric transformation parameter $\lambda_i$, scalar value $a_i$ and positions and timings $(x_i, y_i, t_i)$. Intuitively, this model corresponds to a dense mixing of stereotyped, static textons as in~\citep{Galerne11}.

As was previously mentioned~\citep{Leon12,Vacher15}, this set of Motion Clouds stimuli is equivalently defined by:
\eql{\label{eq-MC}
I (x, y, t) = \sum_{i \in \NN} a_i \cdot (g_{\lambda_i}(\phi(x, y, t)) \ast \delta(x-x_i, y-y_i, t-t_i))
}
where $\ast$ denotes the convolution operator. Noting the Fourier transform as $\mathcal{F}$ and $(f_x, f_y, f_t) \in \RR^2 \times \RR$ the frequency variables, the image $I$ is thus a stationary Gaussian random field of covariance having the power-spectrum $\mathcal{E}=\mathcal{F}(\phi)$ (for a proof, see~\citep{Vacher15}). It comes 
\begin{equation*}
   \mathcal{F}(I) (f_x, f_y, f_t) = A \cdot e^{i \cdot \Phi} \cdot \mathcal{E}  (f_x, f_y, f_t)
\end{equation*}
where
\begin{equation*}
A \cdot e^{i \cdot \Phi} = \sum_{i\in\NN} a_i \cdot e^{-2i\pi (f_x \cdot x_i + f_y \cdot y_i + f_t \cdot t_i)}
\end{equation*}
corresponds to an iid random phase field scaled by $A \in \RR^+$. Such a field was called a Motion Cloud (MC) and it is parameterized by the (positive-, real-valued) envelope $\mathcal{E}$. To match the statistics of natural scenes or some category of textures, the envelope $\mathcal{E}$ is usually defined as some average spatio-temporal coupling over a set of natural images. Note that it is possible to consider any arbitrary texton $\mathcal{F}^{-1}(\mathcal{E})$, which would give rise to more complicated parameterizations for the power spectrum, but we decided here to stick to a simple case which is relevant to study coding in the retina. In particular, we limited ourselves to rotationally symmetric envelopes and to spatio-temporal scale transformations (zooms).
%

The originality provided here is to introduce an explicit sparsity term in the summation. As such, this sparse Motion Clouds is similarly defined by equation~\ref{eq-MC} but with a sparse distribution of coefficients $a_i$. Mathematically, the $a_i$ are drawn from a heavy-tailed probability distribution function $p_a$ such as that found in natural images (See Figure~\ref{fig:retina_sparseness}-C). Such assumptions were previously used for generating procedural noise in computer vision~\citep{Lagae09}, but we focus here on the generation on a model of sparseness in images. Such endeavor was initiated by~\citep{Sallee03} by defining a mixture of a Dirac and a Gaussian distributions, but here, we derive it from the sparse coding observed in a set of natural images. First let's define $E$ the (sparse) event matrix corresponding to the events' scalar values, parameters, positions and timings forming the set of non-zero values $\mathcal{S}$, that is with $E=\sum_{i\in\mathcal{S}} a_i \cdot \delta(x-x_i, y-y_i, t-t_i)$. For the sake of simplicity, the sparseness parameter $\epsilon$ will thus denote the relative ratio of non-zero coefficients from a fully dense matrix with $\epsilon=1$ to an empty matrix with $\epsilon=0$.
First note that this parameter can be trivially linked to the sparseness coefficient that we defined for natural images as  $\epsilon\propto\exp(-\rho)$. 
We also note that we can still compute the image in Fourier space, but with a random phase field such that
\begin{equation*}
A \cdot e^{i \cdot \Phi} = \sum_{i\in\mathcal{S}} a_i \cdot e^{-2i\pi (f_x \cdot x_i + f_y \cdot y_i + f_t \cdot t_i)}
\end{equation*}
Second, we note that the random phase field is simply the Fourier transform on $E$:
\eql{\label{eq-phase}
A \cdot e^{i \cdot \Phi} = \mathcal{F}(E)
}
Compared to the pure summation, a first advantage of this procedure is that the computation time does not depend on the number of events. A further advantage of equation~\ref{eq-phase} in generating these stimuli is that for any given instance of the noise, we know the position (x, y, t) of each event. As they match the shot noise image of drops of rain on water, we coined this set of stimulus as ``DropLets''. The implementation of this texture generation is of the same order of complexity as that of Motion Clouds and allows for the easy control of the sparseness in the stimulus (See Figure~\ref{fig:DropLets}). Once this set of ``DropLets'' stimuli is defined, we may now design a protocol to stimulate the retina and explicitly study the role of sparseness.
\section{Results: neural activity in a population of ganglion cells}
\begin{figure}
\centering{
\includegraphics[width=.99\linewidth]{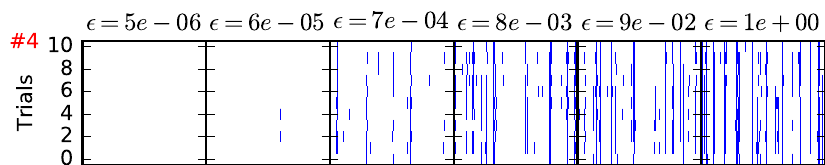}
\includegraphics[width=.99\linewidth]{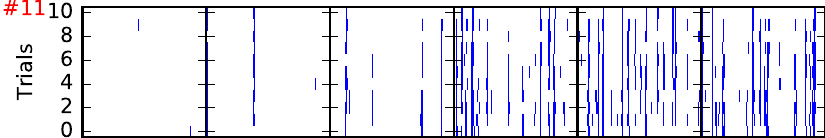}
\includegraphics[width=.99\linewidth]{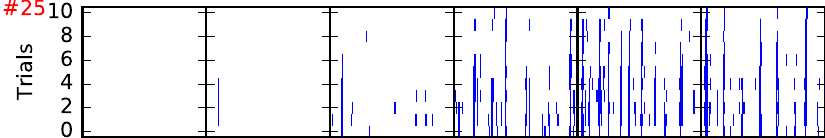}
\includegraphics[width=.99\linewidth]{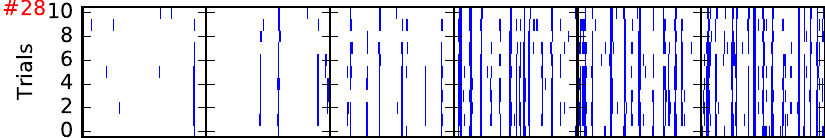}
}
\caption{
{\bf Raster plots for different representative ganglion cells} (different rows, labelled by the red number) as a function of sparseness  $\epsilon$ (different columns, same values of $\epsilon$ as in Figure~\ref{fig:DropLets}). 
In each raster plot, we present spikes as a function of time (x-axis, the duration of a block is $24$ seconds) and for 11 different trials (y-axis).
First, one observes that the global, mean firing rate of each cell increases with sparseness. This was also consistent for the whole population.
Second, the raster plots show that the spiking is very reproducible across trials, showing that this class of random textures could reliably evoke activity in the retina. Moreover, this pattern is very different across cells, indicating that the visual signal is well multiplexed between ganglion cells.
\label{fig:PSTHs}}%
\end{figure}%
To characterize the response to the DropLets stimuli with respect to different sparseness parameters $\epsilon$, we recorded \emph{ex vivo} the electrophysiological activity of a retinal patch using a multi-electrode array (USB-MEA256, Multi Channel Systems MCS GmbH), mounted on an inverted microscope. Stimuli were projected at $60 Hz$ using a conventional DLP projector with custom optics to focus the image onto the photoreceptor layer with a pixel size of $\approx4.7 \mu m$ and an average irradiance of $70 nW/mm^{2}$. The images were presented using a custom software built upon the PsychToolbox library~\citep{Brainard1997we} for MATLAB. Spike sorting was performed using the procedure described in~\citep{Marre:2012hv}.
The protocol consisted of randomly interleaved $24\s$ long, $100\times100$ pixels DropLets stimuli with $6$ levels of sparseness from very sparse ($\epsilon=5.10^{-6}$) to fully dense ($\epsilon=1$), and with parameters chosen on a geometrical scale (see Figure~\ref{fig:DropLets}). Each condition was repeated $11$  times for a total duration of $\approx 1$ hour\if\Supp1 (see SI Section~\ref{sec:protocol} for a full description of the protocol)\fi.
To analyze the activity evoked by the stimuli, first we computed the post-stimulus raster (PSR) for every condition tested and selected the conditions that elicited robust responses, in the sense that the response was strong and similar among trials of the same condition. The conditions that elicited the largest responses correspond to the sequences with the lowest spatial frequencies tested (data not shown). Rasters show that globally, the firing rates increases with the sparseness parameter $\epsilon$. Qualitatively, a surprising result is that for a majority of ganglion cells, the firing pattern is heterogeneous across cells but very reproducible with different repetitions of the same (frozen) random texture (see Figure~\ref{fig:PSTHs}).

\begin{figure}
\centering{
\includegraphics[width=.99\linewidth]{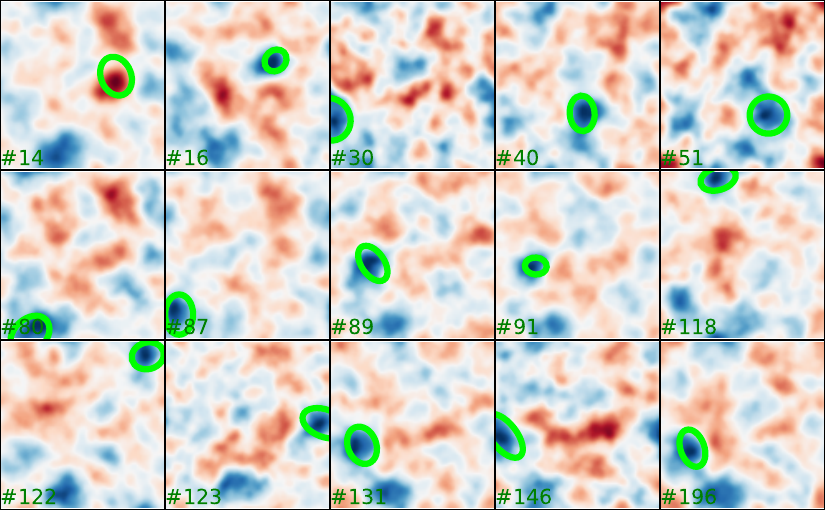}
}
\caption{
{\bf Spike-Triggered Average}: Estimations of the Receptive Fields of representative retinal ganglion cells (whose number is given in green) computed from the response to the droplets (red-blue image) and compared to the fit performed on that computed using a checkerboard stimuli (green ellipse\if\Supp1 ; see Supplementary Figure~\ref{fig:experiment_checkerboard} for a characterization of the whole population\fi). Each image represents the average over the $18$ frames ($\approx25\ms$) preceding a spike. For many cells, our estimation of the RF is very close to the one obtained by the traditional method (albeit generally with a longer temporal profile\if\Supp1 as can be seen in Supplementary Figure~\ref{fig:retina_temp_sta}\fi). Notice that the time of stimulation required is much shorter.
\label{fig:STA}}%
\end{figure}

From the response to these selected sequences, we computed a simple Spike Triggered Average for image matrices of $100\times100$ pixels and 1440 frames for each condition, with a memory of $18$ frames ($\approx25\ms$). Examples of the obtained results can be seen in figure \ref{fig:STA}.
This method yields a good estimation of the RF for some cells, but for many of them the result can be somewhat ambiguous or too noisy, as can be seen in the figure. This can be greatly improved by computing the STA across conditions. As can be seen in figure \ref{fig:STA}, the result is comparable to the estimation obtained with a STA computed from the response from a checkerboard stimulus. However, when analyzing the temporal response, the results are not the equivalent. In the case of the DropLets, the temporal aspect of the RF is always longer or slower than for the RF computed from the checkerboard. In the current work, we did not explore the parameter that controls the lifetime of each droplet, and it appears that the one that was chosen is higher than needed. Another issue, probably a bias caused by the properties of the stimuli, is that for some cells the size of RF is overestimated in comparison with the checkerboard. This is an interesting point that will be further analyzed in future work; if the RF is indeed smaller, then it would mean that the DropLets elicit a response even when also stimulating the periphery of the RF. Another explanation would be that the STA from the checkerboard is underestimating the size of the RF, and in this case the DropLets would yield a better estimation. Further research is already being performed to clarify this issue.

For the moment, we can conclude that it is indeed possible to estimate the RF from the response to some of the DropLets stimuli, however, there are still issues to be solved. First, and the easiest to fix, is the temporality or duration of the features of the stimuli, as apparently they last longer than needed to elicit a response from the cells. The second issue that could be solved is to optimize the size of the features. Clearly, there were a lot of sequences that did not stimulate the retina to get reliable responses, probably because the spatial frequency was too high so the features were undetectable by the retina under the experimental conditions.
Finally, the most important issue would be the noise in the estimation. Even though the RF obtained from the DropLets presented a good match with the ones computed in the traditional way, in many cases the result contained additional signals that may yield conflicting interpretations. One approach to remedy this would be to keep exploring the importance of the sparseness of the stimuli, following the idea that if the features are more isolated in time and space, then there would be less contamination of the STA. Another approach would be to use longer sequences or many sequences with different seeds, to get more spikes triggered by different images so the noise would be reduced when averaging them.
\section{Results: role of sparseness in the efficiency of retinal processing}
The above computation of a STA for this class of stimuli proved that this characterization is not adapted to evaluate the inherent representation corresponding to the measured spike trains. In particular, following previous studies~\citep{Simoncelli2004}, it confirms that the receptive field estimated by the STA varied with the type of stimulus (DropLets versus checkerboard), but also with the sparseness level. As we have shown above (see Figure~\ref{fig:retina_sparseness}), the sparseness level may change drastically in natural images and it is thus reasonable to think that the retina integrates mechanisms acquired during natural evolution to adapt to these different levels of sparseness. In particular, it seems that qualitatively, the spike trains are very reliable across trials for a given sparseness level if it is sufficient to elicit spikes but that at a certain level, there is a larger mixing of different sources (as the cumulative effect of the shot noise model in the texture generation) as we reach a dense mixing of features ($\epsilon=1$). Neural efficiency in the retina is mainly characterized by the capacity of neurons to reliably represent visual information in the spike timings. As such, reliability as measured by the reproducibility of a spike train when being presented the same stimulus is of course not directly related to the efficiency of the neural code, but  provides a convenient information to measure ``post-hoc'' the precision of spike timings with respect to a given stimulus.
\begin{figure}
\centering{\includegraphics[width=\linewidth]{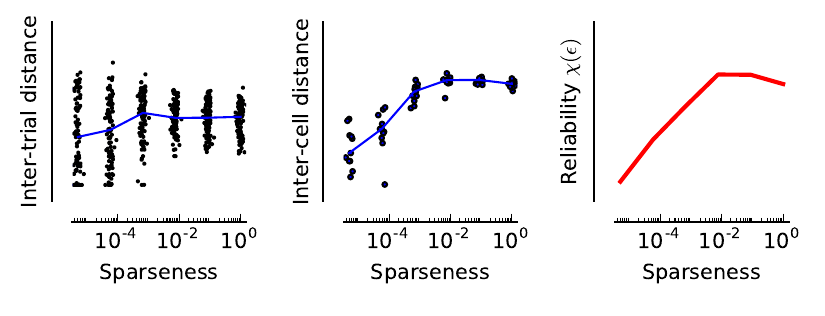}
}
\caption{
{\bf Reliability as a function of sparseness.}
\textsf{(A)}~We first computed for each cell and sparseness level the inter-trial distance as the mean pair distance between two repetitions of the same stimulus, averaged across each block (in time) and for all pairs of trials. We show these measures for each cell as a black dot, and the evolution of the mean for the whole population of ganglion cells (blue line). We show here that this distance decreases for sparseness levels closer to zero: spike trains are more reproducible when the stimulus is sparser.
\textsf{(B)}~As a control, we also measured the inter-cell distance. For any given trial (a blue dot), it is equal to the mean across all pairs of cells of the temporally-averaged distance between the pair of cells. it shows that as $\epsilon$ increases, the mean over trials of the inter-cell distance also increases. 
\textsf{(C)}~Finally, we compute the relative reliability as the ratio between the inter-trial distance to the inter-cell distance. This shows that reliability is optimal for a specific range of $\epsilon$ values.
\label{fig:efficiency}}%
\end{figure}%

To estimate quantitatively such intuition, we measured reliability as the average distance between all pairs of trials for any given condition $\epsilon$ (sparseness) and cell $c$ (over the whole population). For any pair of observed spike trains, this distance was given as the SPIKE-distance as implemented in the PySpike package~\citep{Kreuz2013}. This measure gives for any pair $(j, k)$ a value $S^{j,k}$ between $0$ and $1$, where $0$ is achieved if and only if the spike trains are identical. The average over the $N=11$ repetitions for any given condition, and for any given cell gives an average distance (also between $0$ and $1$) equal to (see Figure~\ref{fig:efficiency}-A):
\begin{equation*}
S^a (\epsilon, c)  = \frac{1}{N(N-1)/2}\sum_{j=1}^{N-1} \sum_{k=j+1}^N S^{j, k} (\epsilon, c)
\end{equation*}
Our goal is to compare this reliability for different levels of sparseness and in particular for different average levels of firing rate. To be invariant to such properties, we used an additional measure by measuring for each trial $k$ the average distance between all pairs of cells (also between $0$ and $1$) equal to (see Figure~\ref{fig:efficiency}-B):
\begin{equation*}
S^c (\epsilon, k)  = \frac{1}{M(M-1)/2}\sum_{j=1}^{M-1} \sum_{k=j+1}^M S^{j, k} (\epsilon, k)
\end{equation*}
where $M=101$ corresponds to the number of cells and the pair of spike trains $\{j, k\}$ is chosen for any given fixed value of $\epsilon$ and $k$. Ultimately, for any given sparseness value $\epsilon$, we define the reliability $\chi(\epsilon)$ as the inverse ratio between the distance average over trials and the global distance over cells :
\begin{equation*}
\chi(\epsilon) = \frac{<S^c(\epsilon, k)>_{k \in 1\ldots N}}{<S^a(\epsilon, c)>_{c \in 1\ldots M}}
\end{equation*}
where $<\cdot_j>_{j}$ denotes  the average of $\cdot_j$ over the different $j$. In Figure~\ref{fig:efficiency}-C, we show a plot of $\chi(\epsilon)$ for the different cells with respect to the sparseness level $\epsilon$. It shows that ratio increases (and thus that reliability increases) for the lowest values of sparseness and then saturates at a given level and finally decreases for a fully dense stimulus.

From these results it appears that in the population of recorded cells, efficiency as inferred by our measure of reliability, varied as a function of sparseness. Indeed, the results show that there is a high variability in the measured z-scores for the different cells, but that the overall trend shows that reliability increases with sparseness until it reaches an optimal value, and then it slightly decreases for the dense texture ($\epsilon=1$). This quantitative measure is consistent with our qualitative intuition, that is, that the efficiency of the representation as measured by this reliability measure is optimal for a value of sparseness which is lower than a dense mixing of feature and that we measure here to $\epsilon=8e-03$. It should be however noted that in this experiment, we sampled only a limited number of sparseness values. Moreover, it is likely that the mechanism is implemented by a neural circuit (notably the circuitry horizontal to the retinal surface) which is most certainly adaptive and would ultimately give a different response when presented over a longer term with stimuli with for instance a lower average sparseness value. Finally, it remains still to be explored if a similar behavior would be observed this time for natural images with varying levels of sparseness.
\section{Discussion}
In this paper, we have characterized the different levels of sparseness which are present in natural images, synthesized random textures with parameterized levels of sparseness and finally, we have shown that the neural code as recorded in ganglion cells \emph{ex vivo} responds differentially to these different levels of sparseness. First, we have replicated the observation that natural images follow a prototypical structure for the probability density function of the coefficients that characterize them. Importantly, we have shown that on an image-by-image basis, this structure is well captured by a single parameter $\rho$ which encompass the sparseness of a given image, from a dense texture to a highly sparse configuration. Based on these results, we designed random texture which replicate this parameterization of their sparseness using the ratio of non-zero coefficients $\epsilon\propto\exp(-\rho)$. We used that set of stimuli to evaluate the response on ganglion cells on an \emph{ex vivo} preparation and analyzed the response of the population of neurons as a function of the sparseness parameter. Such analysis proved that the retinal neural code showed differential response with respect to the sparseness of images and preliminary measures on efficiency suggest that it is tuned for a given level of sparseness.
We have shown that there is a limit in sparseness for which the retina can respond optimally; beyond that point, the response is more or less the same, meaning that the retina is still coding the features present in the sparser sequences but is not responding to the additional features of the more dense sequences. We can relate these results to the hypothesis of optimal coding of natural scenes~\citep{Geisler:2008gu}, in which the visual system has a limited capacity to transmit information, it has adapted through evolution to a small subset of all the possible images and optimally code them, discarding the irrelevant information. Thus, the efficiency of the retina to code the stimuli reaches its peak at a sparseness level that would be closer to what the system has evolved to code, and when presented with stimuli containing more signals, it discards the additional information. Much research has been performed to investigate the relationship between natural images and optimal coding, although the focus has been mainly on the spatiotemporal correlations~\citep{Pitkow:2012dh,Rikhye:2015bg}. We have shown here that when keeping the spatiotemporal components constant, the modulation of the sparseness of the stimuli has an evident effect on the retinal response, and more importantly, it allows us to see the level of sparseness beyond which the efficiency does not increase.

It is important to note that these results are the outcome of an interdisciplinary convergence between image processing (to characterize sparseness in natural images), mathematical modeling (for the synthesis of textures) and neurophysiology (for the recordings and their analysis). In particular, we demonstrated here an original framework in which neural recordings are not analyzed \emph{post-hoc}, but are instead tuned by the design of parameterized stimuli. In particular, these stimuli are defined from the analysis of natural images. One limit of this study is that we limited ourselves to a simplistic class of textons (Mexican-hat shaped profiles\if\Supp1 , see Supplementary Figure~\ref{fig:retina_sparseness_dog}\fi), both for the analysis and synthesis but that ganglion cells in the retina are known to be selective to a wider class of stimulations. However, we believe that this class of stimuli is general enough to characterize a wide range of different cell types. Indeed, by looking at local combinations of events (such as doublets), one could characterize different sub-types of ganglion cells, both static (ON, OFF), oriented or even moving, such as to characterize for instance directionally selective ganglion cells. In particular, by manipulating the statistics in the event's matrix $E$, one could target more specifically each of these sub-types.

Finally, the framework proposed in this paper calls for a novel mathematical characterization of the neural code. Indeed, while the Linear / Non-Linear model underlying the computation of the STA has proven to reliably predict neural responses, this has been at the price of a complex machinery.  We believe that a major limitation of this model is the fact that the spiking mechanism is modeled by an inhomogeneous Poisson process. Indeed, while mathematically tractable, it is however a poor model of the spiking mechanism observed in most cells. In particular, the difference of behavior between the Poisson model and neurons is most prominent when using sparse stimuli such as the one present in the most important set of  stimuli for an animal, that is, natural scenes. Such precise firing has been recently shown in area V1 and may originate from a canonical push-pull mechanism~\citep{Kremkow16}. Similarly, such a mechanism is likely to be present as early as in the retina but novel types of models would be necessary to uncover this aspect of neural computations.
%
\newpage
\if\Supp1

\section{Supplementary material}
\subsection{Sparse coding of natural images}
\label{sec:sparse} 
A method for measuring the statistics of edge co-occurrences
in natural images was demonstrated by~\citet{Geisler01}.
Here we extend their method in two important ways.
First, we use an over-complete, multi-scale representation of edges,
which is more similar to the receptive fields in the retina.
Second, we use a synthesis model for the edge representation,
so that the edges we detect are guaranteed to be sufficient to regenerate the image with a low error.

The first step of our method involves defining the dictionary of templates (or
filters) for detecting edges.
We use a log-Gabor representation, which is well suited to represent
a wide range of natural images~\citep{Fischer07}.
This representation gives a generic model of edges parameterized by their shape,
orientation, and scale.  We set the range of these parameters to match what has been reported
for the responses in rodents' retina.
In particular, we set the bandwidth of the Fourier representation of the filters
to $1$ and $\infty$ respectively in log-frequency and polar coordinates
to get a family of circular but scale-selective filters
(see~\citet{Fischer07cv} and  Supplementary Figure~\ref{fig:retina_sparseness_dog} for examples of such edges).
This architecture is similar to that used by~\citet{Geisler01}.
Prior to the analysis of each image, we used the spectral whitening filter
described by~\citet{Olshausen97} to provide
a good balance of the energy of output coefficients~\citep{Perrinet03ieee,Fischer07}.

A linear convolution model automatically provides a rotation and translation-invariant representation.
Such invariance can be extended to  scalings
by choosing to multiplex these sets of filters at different spatial scales.
Although orthogonal representations are popular for computer vision
due to their computational tractability,
it is desirable in our context that we have a high over-completeness
in the representation to have a detailed measure of the association field.
Ideally, the parameters of edges would vary in a continuous fashion,
to provide relative translation, rotation, and scale invariance.
We chose to have $8$ dyadic levels (that is, doubling the scale at each level)
for the set of $256\times 256$ images.
Tests with a range of different numbers ofscales yielded similar results.
Finally, each image is transformed into a pyramid of coefficients.
This pyramid consists of approximately $4/3\times256^{2}\approx8.7\times10^4$ pixels
multiplexed on $8$ scales, that is,
approximately $.7\times10^6$ coefficients, an over-completeness factor of about $11$.

This transform is linear and can be performed by a simple convolution
repeated for every edge type.
Following~\citet{Fischer07cv}, convolutions were performed in the Fourier (frequency) domain
for computational efficiency.
The Fourier transform allows for a convenient definition of the edge filter characteristics,
and convolution in the spatial domain is equivalent to a simple multiplication in the frequency domain.
By multiplying the envelope of the filter and the Fourier transform of the image,
one may obtain a filtered spectral image that may be converted to
a filtered spatial image using the inverse Fourier transform.
We exploited the fact that by omitting the symmetrical lobe of the envelope of the filter
in the frequency domain,
the output of this procedure gives a complex number whose
real part corresponds to the response to the symmetrical part of the edge,
while the imaginary part corresponds to the asymmetrical part of the edge
(see~\citet{Fischer07cv} for more details).
More generally, the modulus of this complex number gives the energy response
to the edge (comparable to the response of complex cells in area V1),
while its argument gives the exact phase.
Such a representation is implemented 
Python scripts available at \url{XXX anonymous XXX}. 
This property further expands the richness of the representation.

Because this dictionary of edge filters is over-complete, the linear
representation would give a inefficient representation of the
distribution of edges (and thus of edge co-occurrences) due to {\it a
  priori} correlations between coefficients.  Therefore, starting from
this linear representation, we searched for the most sparse
representation.  Minimizing the $\ell_0$ pseudo-norm (the number of
non-zero coefficients) leads to an expensive combinatorial search with
regard to the dimension of the dictionary (it is NP-hard).  As
proposed first by~\citet{Perrinet02sparse}, we may approximate a
solution to this problem using a greedy approach.

In general, a greedy approach is applied when finding the best combination is
difficult to solve globally, but can be solved progressively,
one element at a time.
Applied to our problem, the greedy approach corresponds to first choosing
the single filter $\Phi_i$ that best fits the image along with a suitable coefficient $a_i$,
such that the single source $a_i\Phi_i$ is a good match to the image.
Examining every filter $\Phi_j$, we find the filter $\Phi_i$
with the maximal correlation coefficient, where:
\begin{equation}
i = \mbox{argmax}_j \left( \left\langle \frac{\mathbf{I}}{\| \mathbf{I} \|} , \frac{
\Phi_j}{\| \Phi_j\|} \right\rangle \right),
\label{eq:coco}
\end{equation}
$\langle \cdot,\cdot \rangle$ represents the inner product, and $\| \cdot  \|$
represents the $\ell_2$ (Euclidean) norm. Since filters at a given scale and orientation
are generated by a translation, this operation can be efficiently computed using a convolution,
but we keep this notation for its generality.
The associated coefficient is the scalar projection:
\begin{equation}
a_{i} = \left\langle \mathbf{I} , \frac{ \Phi_{i}}{\| \Phi_{i}\|^2} \right\rangle
\label{eq:proj}
\end{equation}
Second, knowing this choice, the image can be
decomposed as
\begin{equation}
\mathbf{I} = a_{i} \Phi_{i} + \bf{R}
\label{eq:mp0} \end{equation}
where $\bf{R}$ is the residual image.
We then repeat this 2-step process on the residual (that is, with $\mathbf{I} \leftarrow \bf{R}$)
until some stopping criterion is met.
Note also that the norm of the filters has no influence in this algorithm
on the choice function or on the reconstruction error.
For simplicity and without loss of generality,
we will thereafter set the norm of the filters to $1$: $\forall j, \| \Phi_j \| =1$.
Globally, this procedure gives us a sequential algorithm for reconstructing the signal
using the list of sources (filters with coefficients), which greedily optimizes the $\ell_0$ pseudo-norm
(i.e., achieves a relatively sparse representation given the stopping criterion).
The procedure is known as the Matching Pursuit (MP) algorithm~\citep{Mallat93},
which has been shown to generate good approximations for natural images~\citep{Perrinet10shl}.

For this work we made two minor improvements to this method:
First, we took advantage of the response of the filters as complex numbers.
As stated above, the modulus gives a response independent of the phase of the filter,
and this value was used to estimate the best match of the residual image
with the possible dictionary of filters (Matching step).
Then, the phase was extracted as the argument of the corresponding coefficient
and used to feed back onto the image in the Pursuit step.
This modification allows for a phase-independent detection of edges,
and therefore for a richer set of configurations,
while preserving the precision of the representation.

Second, we used a ``smooth'' Pursuit step.
In the original form of the Matching Pursuit algorithm,
the projection of the Matching coefficient is fully removed from the image,
which allows for the optimal decrease of the energy of the residual
and allows for the quickest convergence of the algorithm
with respect to the $\ell_0$ pseudo-norm
(i.e., it rapidly achieves a sparse reconstruction with low error).
However, this efficiency comes at a cost,
because the algorithm may result in non-optimal representations
due to choosing edges sequentially and not globally.
This is often a problem when edges are aligned (e.g. on a smooth contour),
as the different parts will be removed independently, potentially leading
to a residual with gaps in the line.
Our goal here is not to get the fastest decrease of energy,
but rather to provide a good representation of edges along contours.
We therefore used a more conservative approach,
removing only a fraction (denoted by $\alpha$)
of the energy at each pursuit step (for MP, $\alpha=1$).
We found that $\alpha=0.5$ was a good compromise between rapidity and smoothness.
One consequence of using $\alpha<1$ is that, when removing energy along contours,
edges can overlap; even so, the correlation is invariably reduced.
Higher and smaller values of $\alpha$ were also tested,
and gave classification results similar to those presented here.

In summary, the whole learning algorithm is given by the following nested loops
in pseudo-code:
\begin{enumerate}
\item draw a signal $\mathbf{I}$ from the database; its energy is $E = \| \mathbf{I} \|^2$,
\item initialize sparse vector $\mathbf{s}$ to zero and linear coefficients $\forall j, {a}_j=<\mathbf{I}, \Phi_j >$,
\item while the residual energy $E = \| \mathbf{I} \|^2$ is above a given threshold do:
\begin{enumerate}
\item select the best match: $i = \mbox{ArgMax}_{j} | {a}_j |$, where $| \cdot |$ denotes the modulus,
\item increment the sparse coefficient: $s_{i} = s_{i} + \alpha \cdot {a}_{i}$,
\item update residual image: $ \mathbf{I} \leftarrow \mathbf{I} - \alpha \cdot a_{i} \cdot \Phi_{i} $,
\item update residual coefficients: $\forall j, {a}_j \leftarrow {a}_j - \alpha \cdot a_{i} <\Phi_{i} , \Phi_j > $,
\end{enumerate}
\item the final non-zero values of the sparse representation vector
$\mathbf{s}$, give the list of edges representing the
image as the list of couples $(i, s_{i})$, where $i$ represents an edge occurrence
as represented by its position, orientation and scale.
\end{enumerate}
This class of algorithms gives a generic and efficient representation of edges,
as illustrated by the example see Figure~\ref{fig:retina_sparseness}. 
The performance of the algorithm can be measured quantitatively
by reconstructing the image from the list of extracted edges.
Measuring the ratio of extracted energy in the images, $N=9192$ edges were
enough to extract an average of $90\%$ of the energy of $256\times 256$
images on all sets of images. 
with packages NumPy (version 1.6.2) and SciPy (version 0.7.2)~\citep{Oliphant07}
on a cluster of Linux computing nodes.
Visualization was performed using Matplotlib (version 1.1.0)~\citep{Hunter07}.
These python scripts are available at \url{XXX anonymous XXX}. 

\begin{figure}
\centering{\includegraphics[width=\linewidth]{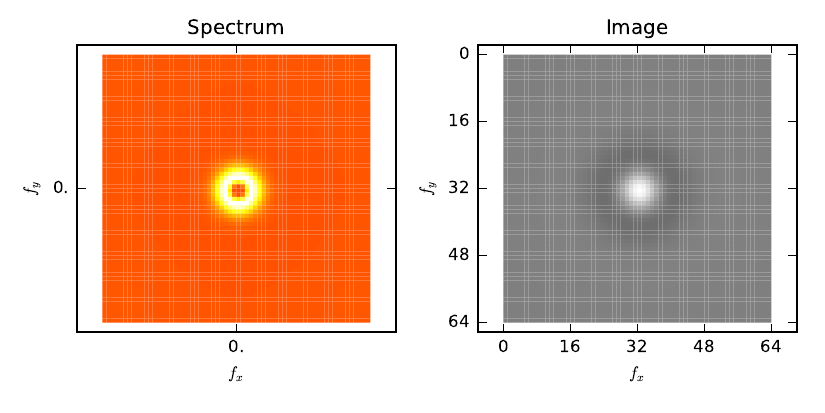}}
\caption{
{\bf Filters used in this paper.} To mimic the average profile of receptive fields in the retina, we use rotationally symmetric filters. These are characterized by a donut-shaped spectrum (Left) and resemble the Laplacian of Gaussian or the Difference of Gaussian filters in Image coordinate (Right)
\label{fig:retina_sparseness_dog}}%
\end{figure}%
\subsection{Defining natural images}
Our goal is to study how the statistics of contrast occurrence vary across a set of natural images.
It consists of the image databases ($600$ images each)\footnote{Publicly available at {\it http://cbcl.mit.edu/software-datasets/serre/SerreOlivaPoggioPNAS07}.} used by~\citet{Serre07},
which contain either animals at different close-up views in a natural setting (which we call ``animal image''),
or natural images without animals, which we call ``non-animal natural images''.
\begin{figure}
\centering{\includegraphics[width=\linewidth]{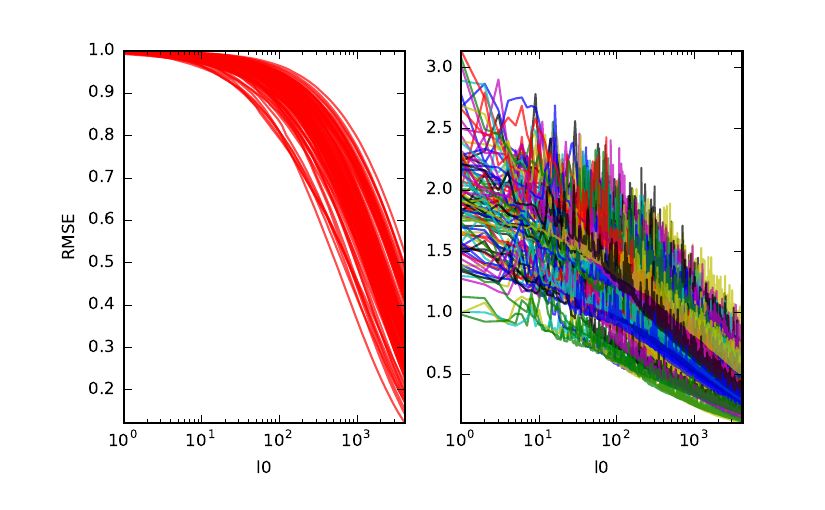}}
\caption{
{\bf Coding of natural images.}
Coefficients decrease as a function of their rank of extraction (average across images $\pm$ standard deviation). This decrease is faster when using a sparse coding mechanism (Matching Pursuit, 'MP') than when just ordering the linear coefficients ('lin'). \textsf{(B-inset)}~For the sparse coding, we controlled the quality of the reconstruction from the edge information such that the residual energy is less than $10\%$ over the whole set of images, a criterion met on average when identifying $8192$ edges per image for images of size $256\times 256$ (that is, a relative sparseness of $\approx 9\%$ of activated coefficients).
\label{fig:retina_sparseness_raw}}%
\end{figure}%
\subsection{DropLets}

\subsection{Designing the protocol}
\label{sec:protocol} 

Designing an experiment is mainly constrained by the time available for recording and we will estimate it here to at maximum 2 hours. The protocol consists in showing a set of motion clouds with different parameters. Spatial frequency will be tuned from the literature for the given eccentricity in retina, while average speed (as in the above example) is always nil. The remaining parameters are the precision in spatial frequency ($B_f$), the precision in time ($B_V$) which is inversely proportional to the average life-time of the textons,  and precision for orientation ($B_\theta = \infty$, orientation $\theta$ is arbitrary).

%
%
\begin{itemize}
\item Presentation of stimuli at 60.0 (frame/second) on the 400x400 (pxl x pxl) array during 24.0 s

\item fixed parameters:
\begin{itemize}
\item  $sf_0 = 0.05$ mean spatial frequency tuned for optimal neural tuning,
\item  $B_sf = 0.075$ is the spatial frequency bandwidth tuned for optimal neural tuning, proportional to $sf_0=0.05$,
\item  $B_V = 0.5$ is the temporal frequency bandwidth tuned for optimal retinal tuning,
\item $B_{\theta} = \infty$ implements circular symmetry, as such, $\theta$ is arbitrary
\end{itemize}
\item parameters:
\begin{itemize}
\item $ N_{sparse} = 6$ degrees of sparseness, distributed on a log-scale of base $sparse_{base}=200000.0$,
         resulting in the sparseness vector $sparseness = [  5.00000000e-06   5.74349177e-05   6.59753955e-04   7.57858283e-03   8.70550563e-02   1.00000000e+00]$ - resulting in the following number of components: $[72, 827, 9500, 109131, 1253592, 14400000]$
\item $N_{trial} = 11$ different repetitions
%
\end{itemize}
\end{itemize}
%
%
\begin{figure}
\centering{
\includegraphics[width=.49\linewidth]{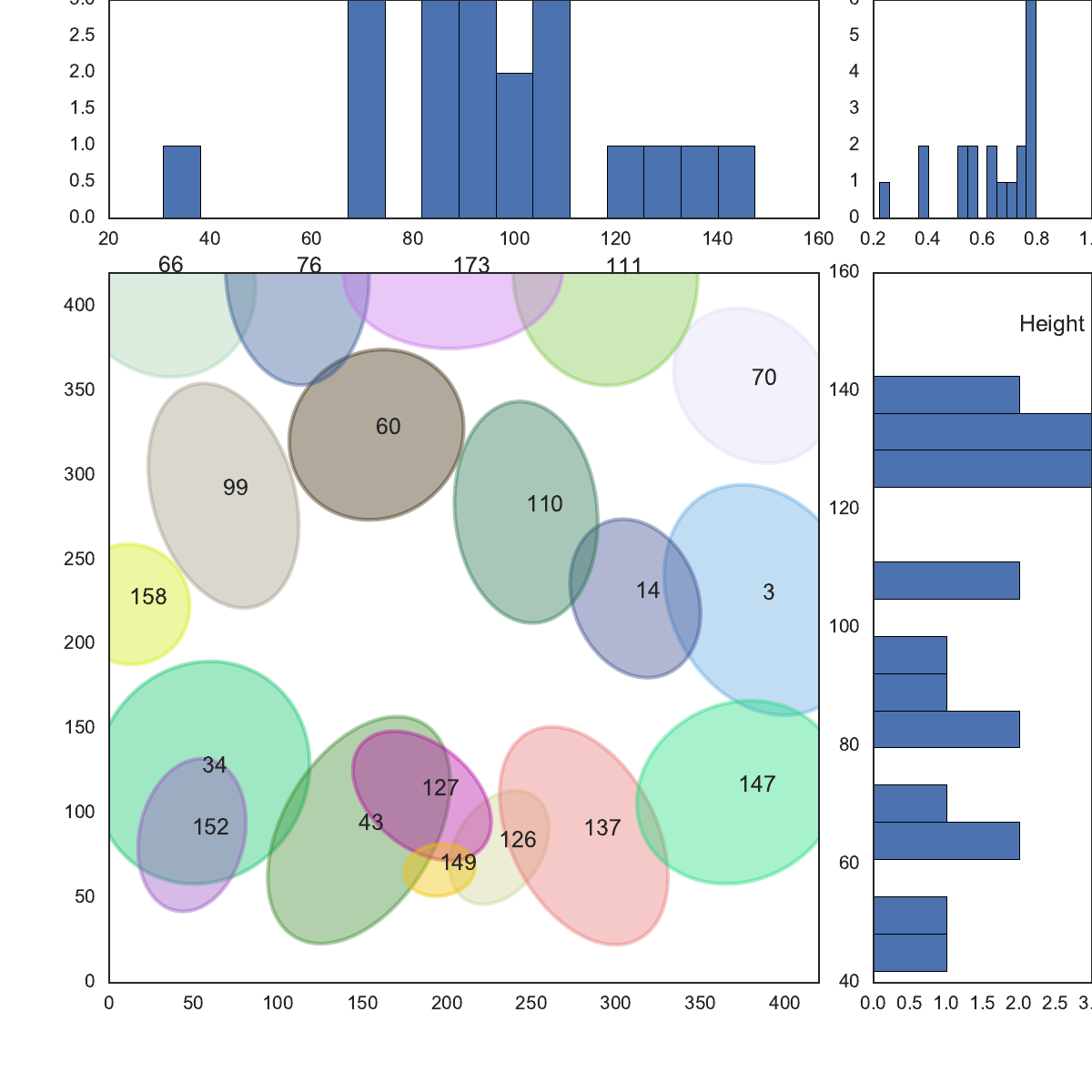}
\includegraphics[width=.49\linewidth]{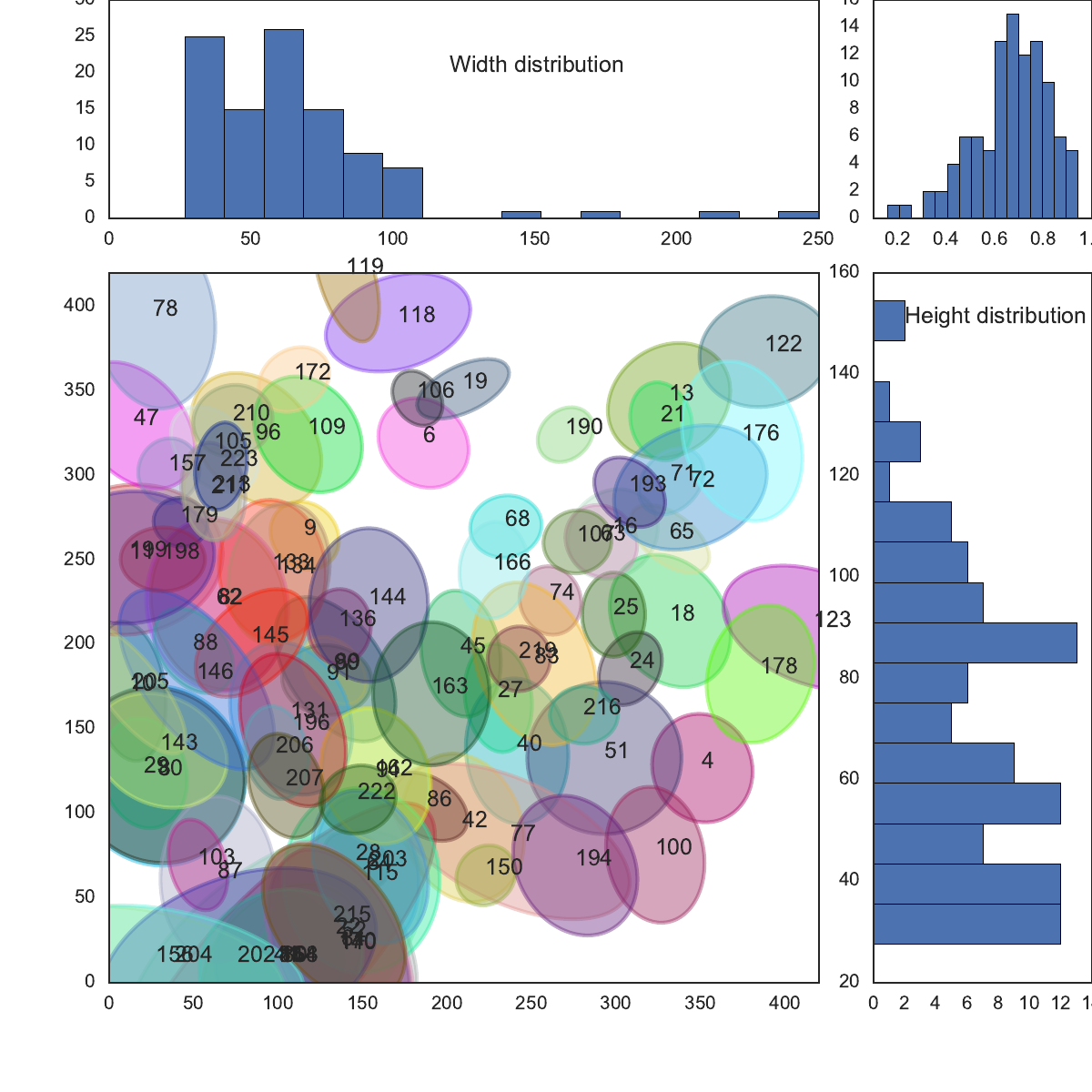}
}
\caption{
{\bf Characterizing cells using the checkerboard stimulus:} Estimations of the Receptive Fields of representative retinal ganglion cells using a checkerboard stimulus.  On the left we show the ON cells and on the right the OFF cells. The large plot is the map of the location of each receptive field. We show the ellipses corresponding to the best fits on the RFs' shapes.
 Adjacent to them are the histograms of the distribution of height and width, and in the corner is the distribution of the eccentricity of the ellipses fitted to the RF. 
\label{fig:experiment_checkerboard}}%
\end{figure}%

\begin{figure}
\centering{
\includegraphics[width=\linewidth]{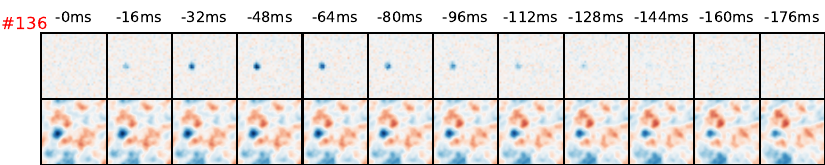}
\includegraphics[width=\linewidth]{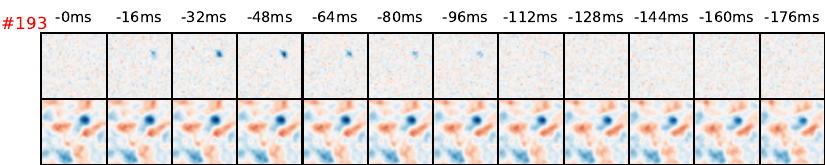}
\includegraphics[width=\linewidth]{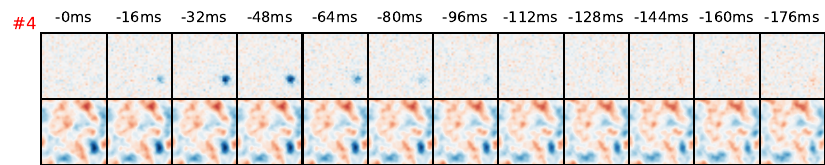}
}
\caption{
{\bf Temporal profiles of the estimated Receptive Fields:}
Each pair of rows represent the STA of a representative cell. Each image is the average frame preceding each spike, at the time indicated above. The upper row is the STA computed from the response to the checkerboard stimulus and the lower row the STA computed from the response to the DropLets. As can be seen, the temporal course is much shorter for the checkerboard, mainly due to the temporal correlations of the DropLets. Ideally, the parameters will be tuned to get an equivalent response from both.
For many cells, our estimation of the RF is very close to the one obtained by the traditional method (albeit generally with a longer temporal profile). Notice that the time of stimulation required is much shorter.
\label{fig:retina_temp_sta}
}%
\end{figure}%

\fi
\section*{References}

\small

%
%
\printbibliography
\end{document}